\documentclass[a4paper,preprint]{revtex4}

\usepackage{amsmath,amssymb,amsfonts}
\usepackage{graphicx}

\begin{document}

\title{On the convergence of Kikuchi's natural iteration method}

\author{Marco Pretti}

\affiliation{Istituto Nazionale per la Fisica della Materia (INFM)
and Dipartimento di Fisica, \\ Politecnico di Torino, Corso Duca
degli Abruzzi 24, I-10129 Torino, Italy}

\date{\today}

\begin{abstract}
In this article we investigate on the convergence of the natural
iteration method, a numerical procedure widely employed in the
statistical mechanics of lattice systems to minimize Kikuchi's
cluster variational free energies. We discuss a sufficient
condition for the convergence, based on the coefficients of the
cluster entropy expansion, depending on the lattice geometry. We
also show that such a condition is satisfied for many lattices
usually studied in applications. Finally, we consider a recently
proposed general method for the minimization of non convex
functionals, showing that the natural iteration method turns out
as a particular case of that method.
\end{abstract}


\maketitle

\section{Introduction}

The cluster variation method (CVM) is a powerful approximate
technique for the statistical mechanics of lattice systems, which
can improve the simple mean field and Bethe theories, by taking
into account correlations on larger and larger distances. It was
first proposed by Kikuchi in 1951~\cite{Kikuchi1951} as an
approximate evaluation of the thermodynamic weight of the system,
and since then it has been reformulated several
times~\cite{Morita1972,Schlijper1983,An1988}, mainly to clarify
the nature of the approximation and to simplify the way to work it
out. Quite a recent formulation~\cite{An1988} shows that the CVM
consists in a truncation of the cumulant entropy expansion. Each
cumulant is associated to a cluster of sites and the truncation is
justified by the expected rapid vanishing of the cumulants upon
increasing the cluster size. In this way the CVM can be viewed as
a hierarchy of approximations, each one defined by the set of
maximal clusters retained in the cumulant expansion, usually
denoted as basic clusters. If pairs of nearest neighbor sites are
chosen as basic clusters, the CVM coincides with the Bethe
approximation. Generally, using larger basic clusters improves the
approximation, even if the convergence of the cumulant expansion
to the exact entropy has been rigorously proved just in a few
cases~\cite{Schlijper1983,Kikuchi1994}.

Due to its relative simplicity and accuracy, the CVM is widely
used in every kind of statistical mechanical applications, to
determine both thermodynamic
properties~\cite{GiaconiaPagotTetot2000,AstaHoyt2000,SchonInden1996}
and phase
diagrams~\cite{Kentzinger2000,Lopez-Sandoval1999,Oates1999,BuzanoPretti1997}.
The CVM results generally compare well with those of Monte Carlo
simulations~\cite{Lopez-Sandoval1999,Oates1999,Lapinskas1998} as
well as experimental
ones~\cite{ClouetNastarSigli2004,SchonInden2001,Kentzinger2000,GiaconiaPagotTetot2000,Lapinskas1998,SchonInden1996}.
Making use of suitable series of CVM approximations, it is also
possible to extrapolate quite accurate estimates of critical
exponents~\cite{Pelizzola2000,Pelizzola1994,KatoriSuzuki1994,KatoriSuzuki1988}.
Recently, it has been shown that the belief propagation algorithm,
an approximate method for statistical inference, employed for a
lot of technologically relevant problems
(image~\cite{TanakaInoueTitterington2003} and signal
processing~\cite{Kschnischang2002}, decoding of error-correcting
codes~\cite{Kschnischang2002,Frey1998}, machine
learning~\cite{Frey1998}), is actually equivalent to the
minimization of a Bethe free energy for statistical mechanical
models defined on graphs~\cite{YedidiaFreemanWeiss2002}. This fact
has opened new research areas both to the application of the CVM
as an improvement of the
approximation~\cite{YedidiaFreemanWeiss2002}, and to the analysis
of efficient minimization
algorithms~\cite{Yuille2001,HeskesAlbersKappen2003,PrettiPelizzola2003},
mainly due to the fact that belief propagation sometimes fails to
converge.

Let us introduce the problem from the CVM point of view. Once the
approximate entropy (and hence free energy) for the chosen set of
basic clusters has been obtained, one has to face the problem of
minimizing a complicated non-convex functional in the basic
cluster probability distributions. An algorithm for minimizing
such a functional has been proposed by Kikuchi
himself~\cite{Kikuchi1974}, and is known as natural iteration
method (NIM). A proof of convergence of this algorithm has been
given in the original paper, essentially for the Bethe
approximation, which can be easily extended to the Husimi
tree~\cite{Pretti2003}. Nevertheless, the range of convergent
cases seems to be much wider, so that the natural iteration method
might be interesting also for the non conventional applications
mentioned above.

In this article we analyze a sufficient condition for the
convergence of the NIM. Such a condition is a requirement on the
coefficients of the cluster entropy expansion (obtained from the
cumulant expansion through a M{\"o}bius inversion~\cite{An1988})
and is shown to hold for quite a large variety of approximations
that are generally used to treat thermodynamic systems. Namely, we
consider: a set of ``plaquette'' approximations on different
lattices~\cite{Kikuchi1974,SchonInden1996,BuzanoPretti1997,KingChen1999},
Kikuchi's B~and C~hierarchies for the
square~\cite{KikuchiBrush1967} and
triangular~\cite{PelizzolaPretti1999} lattices, the cube
approximation for the simple cubic lattice. As far as the latter
case is concerned, we actually analyze a generic hypercube
approximation on the hypercubic lattice in $d$~dimensions, showing
that the sufficient condition holds for $d \leq 3$. Finally we
take into account a recently proposed algorithm for the
minimization of the CVM free energy~\cite{HeskesAlbersKappen2003},
which allows several alternatives, depending on the possibility of
upperbounding the free energy with convex (easy to be minimized)
functions. We show that one of the best choices is actually
equivalent to the natural iteration method.

\section{The CVM free energy}

As mentioned in the Introduction, the approximate CVM entropy can
be written as a linear combination of cluster
entropies~\cite{An1988}
\begin{equation}
  S = \sum_\alpha a_\alpha S_\alpha
  ,
  \label{eq:sumrule}
\end{equation}
where the sum index~$\alpha$ runs over all basic clusters and
their subclusters. We shall always consider clusters in this set
only. The cluster entropies are defined as usual
\begin{equation}
  S_\alpha
  =
  - \sum_{x_\alpha}
  p_\alpha(x_\alpha) \log p_\alpha(x_\alpha)
  ,
\end{equation}
where $p_\alpha(x_\alpha)$ denotes the probability of the
configuration~$x_\alpha$ for the cluster~$\alpha$, the sum runs
over all possible configurations, and the Boltzmann constant~$k$
is set to~$1$ (entropy is measured in natural units). The
coefficients can be determined recursively, starting from basic
clusters down to subclusters, making use of the following
property~\cite{An1988}
\begin{equation}
  \sum_{\alpha' \supseteq \alpha} a_{\alpha'} = 1
  \ \ \forall \alpha
  .
\end{equation}
Due to the fact that a basic cluster~$\gamma$ never contains (by
definition) another basic cluster, from the above formula we
immediately get $a_\gamma = 1 \ \forall \gamma$. Here and in the
following, $\gamma$ denotes basic clusters. As far as the
hamiltonian is concerned, we assume that it can be written as a
sum of contributions $h_\gamma$ from all basic clusters as
\begin{equation}
  \mathcal{H} = \sum_{\gamma} h_\gamma(x_\gamma)
  ,
\end{equation}
where of course $x_\gamma$ denote basic cluster configurations.
Let us decide to write the whole CVM free energy as a sum over
basic clusters, splitting entropy contributions from each
subcluster among all basic clusters that contain it (in equal
parts). Assuming energies normalized to $kT$, we obtain
\begin{equation}
  F[p] =
  \sum_{\gamma}
  \sum_{x_\gamma} p_\gamma(x_\gamma)
  \left[
  h_\gamma(x_\gamma)
  + \log p_\gamma(x_\gamma)
  + \sum_{\alpha \subset \gamma} b_\alpha \log p_\gamma(x_\alpha)
  \right]
  ,
  \label{eq:f1}
\end{equation}
where
\begin{equation}
  p_\gamma(x_\alpha) \equiv
  \sum_{x_{\gamma \setminus \alpha}}
  p_\gamma(x_\gamma)
  .
  \label{eq:margin1}
\end{equation}
Let us notice that we have defined new coefficients $b_\alpha
\equiv a_\alpha / c_\alpha$, where $c_\alpha$ denotes the number
of basic clusters that contain~$\alpha$, and we have expressed
subcluster probability distributions as marginals of basic cluster
distributions, according to Eq.~\eqref{eq:margin1} (the sum runs
over configurations $x_{\gamma \setminus \alpha}$ of the basic
cluster $\gamma$ minus the subcluster $\alpha$).

\section{The natural iteration method}

In the above formulation, basic cluster distributions
$\{p_\gamma(x_\gamma)\}$ are the variational parameters of the
free energy (which is denoted in short by $F[p]$), and the
thermodynamic equilibrium state can be determined by minimization
with respect to these parameters with suitable normalization and
compatibility constraints. By compatibility we mean of course that
marginal distributions $p_\gamma(x_\alpha)$ must be the same for
all basic clusters $\gamma \supset \alpha$. Let us notice that,
for most thermodynamic applications, one usually makes some
homogeneity assumption on the system, and this generally reduces
the problem to only one or few different basic cluster
distributions. Compatibility constraints may be still necessary to
impose the required symmetry. We go on with the complete
formulation, without loss of generality. The important thing is
that in any case we deal with constraints that are linear in the
probability distributions (compatibility), possibly with an
additive constant (unit) term (normalization). According to the
Lagrange method, we transform the constrained minimum problem with
respect to $\{p_\gamma(x_\gamma)\}$ to a free minimum problem for
an extended functional which depends on additional parameters
(Lagrange multipliers). Due to linearity, the extended functional
can be written in the form
\begin{equation}
  \tilde{F}[p,\lambda] = F[p]
  - \sum_{\gamma} \sum_{x_\gamma} p_\gamma(x_\gamma) \lambda_\gamma(x_\gamma)
  ,
  \label{eq:f2}
\end{equation}
where $\{\lambda_\gamma(x_\gamma)\}$ are the Lagrange multipliers.
Of course, $\{\lambda_\gamma(x_\gamma)\}$ are not all independent
variables, but internal relationships are system dependent, and we
do not analyze them. Let us only notice, for future use, that the
difference between the new functional and the original one (the
last term in Eq.~\eqref{eq:f2}) is actually independent of the
$\{p_\gamma(x_\gamma)\}$ distributions, provided they satisfy the
required constraints.

The derivatives of~$\tilde{F}$ with respect
to~$p_\gamma(x_\gamma)$ turn out to be
\begin{equation}
  \frac{\partial \tilde{F}[p,\lambda]}{\partial p_\gamma(x_\gamma)}
  =
  h_\gamma(x_\gamma)
  + \log p_\gamma(x_\gamma)
  + \sum_{\alpha \subset \gamma} b_\alpha \log p_\gamma(x_\alpha)
  - \lambda_\gamma(x_\gamma)
  + \text{const.}
  ,
\end{equation}
where the additive constant is irrelevant and we can absorb it
into the Lagrange multipliers. Setting the above derivatives to
zero resolves stationarization with respect to probability
distributions. The natural iteration method consists in rewriting
such equations in a fixed point form, that is
\begin{equation}
  \hat{p}_\gamma(x_\gamma) =
  e^{\lambda_\gamma(x_\gamma) - h_\gamma(x_\gamma)}
  \prod_{\alpha \subset \gamma} \left[ p_\gamma(x_\alpha) \right]^{-b_\alpha}
  ,
  \label{eq:nim}
\end{equation}
and then solving them by simple iteration. A new estimate of the
basic cluster probability distribution $\hat{p}_\gamma(x_\gamma)$
is obtained from the previous one $p_\gamma(x_\gamma)$ trough its
marginals $p_\gamma(x_\alpha)$. The Lagrange multipliers must be
determined at each iteration, so that also
$\hat{p}_\gamma(x_\gamma)$ satisfies the required constraints.
This job can be done in different ways by a nested procedure
(inner loop), for instance a Newton-Raphson method or a suitable
fixed point method~\cite{Kikuchi1976,PelizzolaPretti1999}. In this
paper we do not deal with the determination of Lagrange
multipliers, but we only focus on the convergence of the main
loop.

\section{Sufficient condition for the convergence}

As usual for iterative algorithms designed to minimize functionals
that are bounded from below, a proof of convergence can be given
by the decreasing of the functional value at each iteration. This
is actually the case for the natural iteration method. Let us
consider the free energy difference $F[\hat{p}]-F[p]$ for two
subsequent iterations $p,\hat{p}$, where $F[p]$ is defined by
Eqs.~\eqref{eq:f1} and~\eqref{eq:margin1}. Taking the logarithm of
both sides of Eq.~\eqref{eq:nim}, we can rewrite the NIM equations
in two different ways, that are
\begin{equation}
  \log \hat{p}_\gamma(x_\gamma) =
  \lambda_\gamma(x_\gamma) - h_\gamma(x_\gamma)
  - \sum_{\alpha \subset \gamma} b_\alpha \log p_\gamma(x_\alpha)
\end{equation}
\begin{equation}
  \sum_{\alpha \subset \gamma} b_\alpha \log p_\gamma(x_\alpha)
  = \lambda_\gamma(x_\gamma) - h_\gamma(x_\gamma)
  - \log \hat{p}_\gamma(x_\gamma)
  .
\end{equation}
Let us replace the former into $F[\hat{p}]$ and the latter into
$F[p]$. Remembering that probability distributions satisfy the
constraints, whence latter term on the right hand side of
Eq.~\eqref{eq:f2} depends on Lagrange multipliers only, we obtain
\begin{equation}
  F[\hat{p}]-F[p]
  = \sum_\gamma \sum_{x_\gamma}
  \left\{
  p_\gamma(x_\gamma)
  \log \frac{\hat{p}_\gamma(x_\gamma)}{p_\gamma(x_\gamma)}
  - \hat{p}_\gamma(x_\gamma) \sum_{\alpha \subset \gamma}
  b_\alpha \log \frac{p_\gamma(x_\alpha)}{\hat{p}_\gamma(x_\alpha)}
  \right\}
  .
  \label{eq:deltaf2}
\end{equation}
Let us consider the inequality $\log \xi \le \xi-1$, observing
that equality holds if and only if $\xi=1$. By applying this
inequality to the first logarithm (the one involving basic cluster
probability distributions) in Eq.~\eqref{eq:deltaf2}, and taking
into account that distributions are normalized, we obtain
\begin{equation}
  F[\hat{p}]-F[p]
  \leq
  - \sum_\gamma \sum_{x_\gamma}
  \hat{p}_\gamma(x_\gamma) \sum_{\alpha \subset \gamma}
  b_\alpha \log \frac{p_\gamma(x_\alpha)}{\hat{p}_\gamma(x_\alpha)}
  ,
  \label{eq:deltaf3}
\end{equation}
where equality holds if and only if $\hat{p}_\gamma(x_\gamma) =
p_\gamma(x_\gamma) \ \forall \gamma,x_\gamma$. The same result
could be obtained by observing that actually the upperbounded
terms coincide with (minus) the Kullbach-Liebler distances between
the probability distributions $p_\gamma(x_\gamma)$ and
$\hat{p}_\gamma(x_\gamma)$. If all subcluster coefficients
$b_\alpha$ were negative, we could apply the same argument to all
terms, and the upperbound would be zero. Such a situation occurs
for instance in the Bethe~\cite{Kikuchi1974} and Husimi
tree~\cite{Pretti2003} approximations, and the proof of
convergence would be complete. In a general case we have to
require a condition on the $b_\alpha$~coefficients. The basic idea
is to ``couple'' smaller cluster terms with a positive coefficient
to larger cluster terms with a negative coefficient, yielding a
sum of ``negative'' Kullbach-Liebler distances (some between
conditional probability distributions), which can then be
upperbounded by zero. The details are given in the following.

\noindent {\bf Theorem (sufficient condition for the
convergence):} Let $\{b_{\alpha^-|\alpha^+}\}$ be a set of non
negative coefficients (allocation coefficients), defined for each
pair of subclusters $\alpha^-,\alpha^+$, such that
$b_{\alpha^-}<0$, $b_{\alpha^+}>0$, and $\alpha^- \supset
\alpha^+$. If the following properties hold for all basic
clusters~$\gamma$
\begin{eqnarray}
  b_{\alpha^+} =
  & \displaystyle \sum_{\alpha^+ \subset \alpha^- \subset \gamma}
  & b_{\alpha^-|\alpha^+}
  \ \ \ \ \ \ \forall \alpha^+ \subset \gamma
  \label{eq:suffcondplus} \\
  -b_{\alpha^-} \geq
  & \displaystyle \sum_{\alpha^+ \subset \alpha^-}
  & b_{\alpha^-|\alpha^+}
  \ \ \ \ \ \ \forall \alpha^- \subset \gamma
  ,
  \label{eq:suffcondminus}
\end{eqnarray}
then
\begin{eqnarray}
  &&
  F[\hat{p}]-F[p] \le 0
  \label{eq:deltafle0} \\
  &&
  F[\hat{p}]-F[p] = 0 \ \ \Longleftrightarrow \ \
  \hat{p} = p
  .
  \label{eq:deltafeq0iff}
\end{eqnarray}
Eq.~(\ref{eq:deltafle0}) means that the free energy can be
decreasing or constant during the procedure, while
Eq.~(\ref{eq:deltafeq0iff}) assures that it is constant only if
the procedure has already reached convergence (i.e., the free
energy  can only decrease during the procedure). A relevant
consequence of Eq.~\eqref{eq:deltafeq0iff} is that it prevents the
dynamical system defined by the NIM equations from having limit
cycles at constant free energy, which could occur in principle.

\noindent {\bf Proof:} Let us consider the right hand side of
Eq.~\eqref{eq:deltaf3} and split the sum over subclusters $\alpha
\subset \gamma$ in two sums over subclusters $\alpha^+,\alpha^-$
with positive or negative coefficients respectively. Positive
coefficients $b_{\alpha^+}$ can be replaced by
Eq.~\eqref{eq:suffcondplus}, while, according to
Eq.~\eqref{eq:suffcondminus}, negative coefficients can be
replaced by
\begin{equation}
  b_{\alpha^-} =
  - \sum_{\alpha^+ \subset \alpha^-}
  b_{\alpha^-|\alpha^+}
  - d_{\alpha^-}
  ,
\end{equation}
for certain $d_{\alpha^-} \geq 0$. Defining, for each $\alpha^-
\supset \alpha^+$, the conditional probability distributions
\begin{equation}
  p_\gamma(x_{\alpha^-}|x_{\alpha^+})
  \equiv
  \frac{p_\gamma(x_{\alpha^-})}{p_\gamma(x_{\alpha^+})}
  ,
\end{equation}
after some simple manipulations we obtain
\begin{equation}
  - \sum_{\alpha \subset \gamma}
  b_\alpha \log \frac{p_\gamma(x_\alpha)}{\hat{p}_\gamma(x_\alpha)}
  = \sum_{\alpha^- \subset \gamma}
  \left[
  d_{\alpha^-}
  \log \frac{p_\gamma(x_{\alpha^-})}{\hat{p}_\gamma(x_{\alpha^-})}
  + \sum_{\alpha^+ \subset \alpha^-}
  b_{\alpha^-|\alpha^+}
  \log \frac{p_\gamma(x_{\alpha^-}|x_{\alpha^+})}{\hat{p}_\gamma(x_{\alpha^-}|x_{\alpha^+})}
  \right]
  .
\end{equation}
The logarithm inequality $\log \xi \le \xi-1$ can now be applied
to all terms in the previous equation, because all coefficients
are positive (or equivalently we get a sum of Kullbach-Liebler
terms), and the zero upperbound of Eq.~\eqref{eq:deltafle0} is
obtained. As previously mentioned, Eq.~\eqref{eq:deltafeq0iff} is
proved by the fact that the logarithm inequality holds if and only
if $\xi = 1$, i.e., the Kullbach-Liebler distance between two
probability distributions is zero if and only if the two
distributions are equal.~$\blacksquare$

\section{Some particular cases}

In this section we consider some particular choices of basic
clusters, that is, some particular CVM approximations for regular
lattices on which several model systems are defined.

\subsection{``Plaquette'' approximations}

By ``plaquette'' approximations we mean a class of approximations
in which basic clusters are of a unique type (which we denote as
plaquette, for example a square on a square lattice), while
subclusters with non zero coefficients are only single sites and
nearest neighbor pairs. Let us denote such clusters by $1$ and $2$
respectively, and, according to the notation introduced in
Sec.~II, let us denote by $a_1$ and $a_2$ the coefficients of the
cluster entropy expansion, by $c_1$ and $c_2$ the numbers of
plaquettes sharing a given subcluster, and by $b_i = a_i/c_i$ the
normalized coefficients. In this class of approximations, it is
possible to show that all the coefficients can be obtained as a
function of $c_1,c_2$ and of the lattice coordination number~$q$.
Making use of Eq.~\eqref{eq:sumrule}, and remembering that basic
clusters (plaquettes) have unit $a$-coefficient, we can write
\begin{eqnarray}
  &&
  a_2 + c_2 = 1
  \\
  &&
  a_1 + qa_2 + c_1 = 1
  ,
\end{eqnarray}
from which $b_i = a_i/c_i$ are easily obtained:
\begin{eqnarray}
  b_2 & = & -\frac{c_2-1}{c_2}
  \label{eq:b2plaq} \\
  b_1 & = & \frac{q(c_2-1)-(c_1-1)}{c_1}
  .
\end{eqnarray}
Then, we have to impose the sufficient conditions on the
coefficients, Eqs.~\eqref{eq:suffcondplus}
and~\eqref{eq:suffcondminus}. From Eq.~\eqref{eq:b2plaq} we easily
see that~$b_2 \leq 0$, which is ok for upperbounding, but
usually~$b_1 \geq 0$. We then have to couple each site to pairs
that contain it and are contained in a given plaquette. Let us
adopt the strategy of splitting the site coefficient among such
pairs in equal parts, so that, being $b_{2|1}$ the only allocation
coefficient and $r$ the number of pairs,
Eqs.~\eqref{eq:suffcondplus} and~\eqref{eq:suffcondminus} read
\begin{eqnarray}
  b_1 & = & r b_{2|1}
  \\
  -b_2 & \geq & 2 b_{2|1}
  .
\end{eqnarray}
The allocation coefficient may be easily eliminated, yielding the
single condition
\begin{equation}
  \frac{b_1}{r} + \frac{b_2}{2} \leq 0
  .
  \label{eq:condsuffplaq}
\end{equation}
It is possible to show that also the $r$~parameter depends on
$c_1,c_2,q$ only. Let us imagine to multiply the number~$q$ of
nearest neighbor pairs sharing a site times the number $c_2$ of
plaquettes sharing a pair. It is easy to realize that in this way
we have {\em overcounted} $r$~times the number $c_1$ of plaquettes
sharing the given site, i.e.,
\begin{equation}
  rc_1 = qc_2
  .
\end{equation}
With the above manipulation, the condition~\eqref{eq:condsuffplaq}
can be rewritten as
\begin{equation}
  q(c_2-1) \leq 2(c_1-1)
  .
\end{equation}
In this form we can easily verify its validity, which is done in
Tab.~\ref{tab:coefficients} for a set of typical plaquette
approximations. We have considered: the 2d square, triangular, and
honeycomb lattices with a 4-site
square~\cite{BuzanoPretti1997,KingChen1999}, a 3-site
triangle~\cite{KingChen1999}, and an elementary hexagon as basic
cluster respectively, the simple cubic (sc) lattice with a 4-site
square~\cite{KingChen1999} as basic cluster, and the face-centered
cubic (fcc) lattice with a 3-site triangle~\cite{KingChen1999} or
a 4-site tetrahedron~\cite{Kikuchi1974,SchonInden1996} as basic
cluster.

\subsection{B and C hierarchies}

\begin{figure}[p]

  \setlength{\unitlength}{1.2mm}

  \begin{picture}(150,140)(-10,-140)

  \thicklines

  \put(-3,-12){\makebox(15,2)[lb]{\sf BASIC CLUSTER}}
  \put(52,-12){\makebox(15,2)[lb]{\sf PLAQUETTE}}

  \multiput(0,-20)(9,0){5}{\circle*{2}}
  \multiput(4.5,-27.5)(9,0){4}{\circle*{2}}
  \put(0,-20){\line(1,0){9}}
  \put(9,-20){\line(1,0){9}}
  \put(18,-20){\line(1,0){3.5}}
  \put(27,-20){\line(-1,0){3.5}}
  \put(36,-20){\line(-1,0){9}}
  \put(4.5,-27.5){\line(1,0){9}}
  \put(13.5,-27.5){\line(1,0){9}}
  \put(22.5,-27.5){\line(1,0){3.5}}
  \put(31.5,-27.5){\line(-1,0){3.5}}
  \multiput(0,-20)(9,0){4}{\line(3,-5){4.5}}
  \put(4.5,-27.5){\line(3,5){4.5}}
  \put(13.5,-27.5){\line(3,5){4.5}}
  \put(22.5,-27.5){\line(3,5){1.8}}
  \put(27,-20){\line(-3,-5){1.8}}
  \put(31.5,-27.5){\line(3,5){4.5}}
  \put(-1,-18){\makebox(2,2)[lb]{$1$}}
  \put(8,-18){\makebox(2,2)[lb]{$3$}}
  \put(17,-18){\makebox(2,2)[lb]{$5$}}
  \put(19.5,-18){\makebox(3,2)[lb]{$\dots$}}
  \put(24,-18){\makebox(5,2)[lb]{$2L-1$}}
  \put(35,-18){\makebox(5,2)[lb]{$2L+1$}}
  \put(3.5,-31.5){\makebox(2,2)[lt]{$2$}}
  \put(12.5,-31.5){\makebox(2,2)[lt]{$4$}}
  \put(21.5,-31.5){\makebox(2,2)[lt]{$6$}}
  \put(25,-33){\makebox(3,2)[lt]{$\dots$}}
  \put(30.5,-31.5){\makebox(3,2)[lt]{$2L$}}

  \multiput(55,-20)(9,0){2}{\circle*{2}}
  \put(59.5,-27.5){\circle*{2}}
  \put(55,-20){\line(1,0){9}}
  \multiput(55,-20)(9,0){1}{\line(3,-5){4.5}}
  \put(59.5,-27.5){\line(3,5){4.5}}
  \put(54,-18){\makebox(2,2)[lb]{$1$}}
  \put(63,-18){\makebox(2,2)[lb]{$3$}}
  \put(58.5,-31.5){\makebox(2,2)[lt]{$2$}}

  \put(-3,-42){\makebox(15,2)[lb]{\sf PAIR-LIKE CLUSTERS}}
  \put(52,-42){\makebox(15,2)[lb]{\sf PAIRS}}

  \multiput(0,-50)(9,0){4}{\circle*{2}}
  \multiput(4.5,-57.5)(9,0){4}{\circle*{2}}
  \put(0,-50){\line(1,0){9}}
  \put(9,-50){\line(1,0){9}}
  \put(18,-50){\line(1,0){3.5}}
  \put(27,-50){\line(-1,0){3.5}}
  \put(4.5,-57.5){\line(1,0){9}}
  \put(13.5,-57.5){\line(1,0){9}}
  \put(22.5,-57.5){\line(1,0){3.5}}
  \put(31.5,-57.5){\line(-1,0){3.5}}
  \multiput(0,-50)(9,0){4}{\line(3,-5){4.5}}
  \put(4.5,-57.5){\line(3,5){4.5}}
  \put(13.5,-57.5){\line(3,5){4.5}}
  \put(22.5,-57.5){\line(3,5){1.8}}
  \put(27,-50){\line(-3,-5){1.8}}
  \put(-1,-48){\makebox(2,2)[lb]{$1$}}
  \put(8,-48){\makebox(2,2)[lb]{$3$}}
  \put(17,-48){\makebox(2,2)[lb]{$5$}}
  \put(19.5,-48){\makebox(3,2)[lb]{$\dots$}}
  \put(24,-48){\makebox(5,2)[lb]{$2L-1$}}
  \put(3.5,-61.5){\makebox(2,2)[lt]{$2$}}
  \put(12.5,-61.5){\makebox(2,2)[lt]{$4$}}
  \put(21.5,-61.5){\makebox(2,2)[lt]{$6$}}
  \put(25,-63){\makebox(3,2)[lt]{$\dots$}}
  \put(30.5,-61.5){\makebox(3,2)[lt]{$2L$}}

  \put(55,-50){\circle*{2}}
  \put(59.5,-57.5){\circle*{2}}
  \put(55,-50){\line(3,-5){4.5}}
  \put(54,-48){\makebox(2,2)[lb]{$1$}}
  \put(58.5,-61.5){\makebox(2,2)[lt]{$2$}}

  \multiput(9,-70)(9,0){4}{\circle*{2}}
  \multiput(4.5,-77.5)(9,0){4}{\circle*{2}}
  \put(9,-70){\line(1,0){9}}
  \put(18,-70){\line(1,0){3.5}}
  \put(27,-70){\line(-1,0){3.5}}
  \put(36,-70){\line(-1,0){9}}
  \put(4.5,-77.5){\line(1,0){9}}
  \put(13.5,-77.5){\line(1,0){9}}
  \put(22.5,-77.5){\line(1,0){3.5}}
  \put(31.5,-77.5){\line(-1,0){3.5}}
  \multiput(9,-70)(9,0){3}{\line(3,-5){4.5}}
  \put(4.5,-77.5){\line(3,5){4.5}}
  \put(13.5,-77.5){\line(3,5){4.5}}
  \put(22.5,-77.5){\line(3,5){1.8}}
  \put(27,-70){\line(-3,-5){1.8}}
  \put(31.5,-77.5){\line(3,5){4.5}}
  \put(8,-68){\makebox(2,2)[lb]{$3$}}
  \put(17,-68){\makebox(2,2)[lb]{$5$}}
  \put(19.5,-68){\makebox(3,2)[lb]{$\dots$}}
  \put(24,-68){\makebox(5,2)[lb]{$2L-1$}}
  \put(35,-68){\makebox(5,2)[lb]{$2L+1$}}
  \put(3.5,-81.5){\makebox(2,2)[lt]{$2$}}
  \put(12.5,-81.5){\makebox(2,2)[lt]{$4$}}
  \put(21.5,-81.5){\makebox(2,2)[lt]{$6$}}
  \put(25,-83){\makebox(3,2)[lt]{$\dots$}}
  \put(30.5,-81.5){\makebox(3,2)[lt]{$2L$}}

  \put(64,-70){\circle*{2}}
  \put(59.5,-77.5){\circle*{2}}
  \put(59.5,-77.5){\line(3,5){4.5}}
  \put(63,-68){\makebox(2,2)[lb]{$3$}}
  \put(58.5,-81.5){\makebox(2,2)[lt]{$2$}}

  \multiput(0,-90)(9,0){5}{\circle*{2}}
  \put(0,-90){\line(1,0){9}}
  \put(9,-90){\line(1,0){9}}
  \put(18,-90){\line(1,0){3.5}}
  \put(27,-90){\line(-1,0){3.5}}
  \put(36,-90){\line(-1,0){9}}
  \put(-1,-88){\makebox(2,2)[lb]{$1$}}
  \put(8,-88){\makebox(2,2)[lb]{$3$}}
  \put(17,-88){\makebox(2,2)[lb]{$5$}}
  \put(19.5,-88){\makebox(3,2)[lb]{$\dots$}}
  \put(24,-88){\makebox(5,2)[lb]{$2L-1$}}
  \put(35,-88){\makebox(5,2)[lb]{$2L+1$}}

  \multiput(55,-90)(9,0){2}{\circle*{2}}
  \put(55,-90){\line(1,0){9}}
  \put(54,-88){\makebox(2,2)[lb]{$1$}}
  \put(63,-88){\makebox(2,2)[lb]{$3$}}

  \put(-3,-102){\makebox(15,2)[lb]{\sf SITE-LIKE CLUSTERS}}
  \put(52,-102){\makebox(15,2)[lb]{\sf SITES}}

  \multiput(0,-110)(9,0){4}{\circle*{2}}
  \put(0,-110){\line(1,0){9}}
  \put(9,-110){\line(1,0){9}}
  \put(18,-110){\line(1,0){3.5}}
  \put(27,-110){\line(-1,0){3.5}}
  \put(-1,-108){\makebox(2,2)[lb]{$1$}}
  \put(8,-108){\makebox(2,2)[lb]{$3$}}
  \put(17,-108){\makebox(2,2)[lb]{$5$}}
  \put(19.5,-108){\makebox(3,2)[lb]{$\dots$}}
  \put(24,-108){\makebox(5,2)[lb]{$2L-1$}}

  \put(55,-110){\circle*{2}}
  \put(54,-108){\makebox(2,2)[lb]{$1$}}

  \multiput(4.5,-120)(9,0){4}{\circle*{2}}
  \put(4.5,-120){\line(1,0){9}}
  \put(13.5,-120){\line(1,0){9}}
  \put(22.5,-120){\line(1,0){3.5}}
  \put(31.5,-120){\line(-1,0){3.5}}
  \put(3.5,-118){\makebox(2,2)[lb]{$2$}}
  \put(12.5,-118){\makebox(2,2)[lb]{$4$}}
  \put(21.5,-118){\makebox(2,2)[lb]{$6$}}
  \put(25.5,-118){\makebox(3,2)[lb]{$\dots$}}
  \put(30.5,-118){\makebox(2,2)[lb]{$2L$}}

  \put(59.5,-120){\circle*{2}}
  \put(58.5,-118){\makebox(2,2)[lb]{$2$}}

  \multiput(9,-130)(9,0){4}{\circle*{2}}
  \put(9,-130){\line(1,0){9}}
  \put(18,-130){\line(1,0){3.5}}
  \put(27,-130){\line(-1,0){3.5}}
  \put(36,-130){\line(-1,0){9}}
  \put(8,-128){\makebox(2,2)[lb]{$3$}}
  \put(17,-128){\makebox(2,2)[lb]{$5$}}
  \put(19.5,-128){\makebox(3,2)[lb]{$\dots$}}
  \put(24,-128){\makebox(5,2)[lb]{$2L-1$}}
  \put(35,-128){\makebox(5,2)[lb]{$2L+1$}}

  \put(64,-130){\circle*{2}}
  \put(63,-128){\makebox(2,2)[lb]{$3$}}

  \end{picture}

  \caption{
    Basic cluster and subclusters for the B~hierarchy (left side) and
    for the corresponding (triangle) plaquette approximation.
  }
  \label{fig:gerb}
\end{figure}
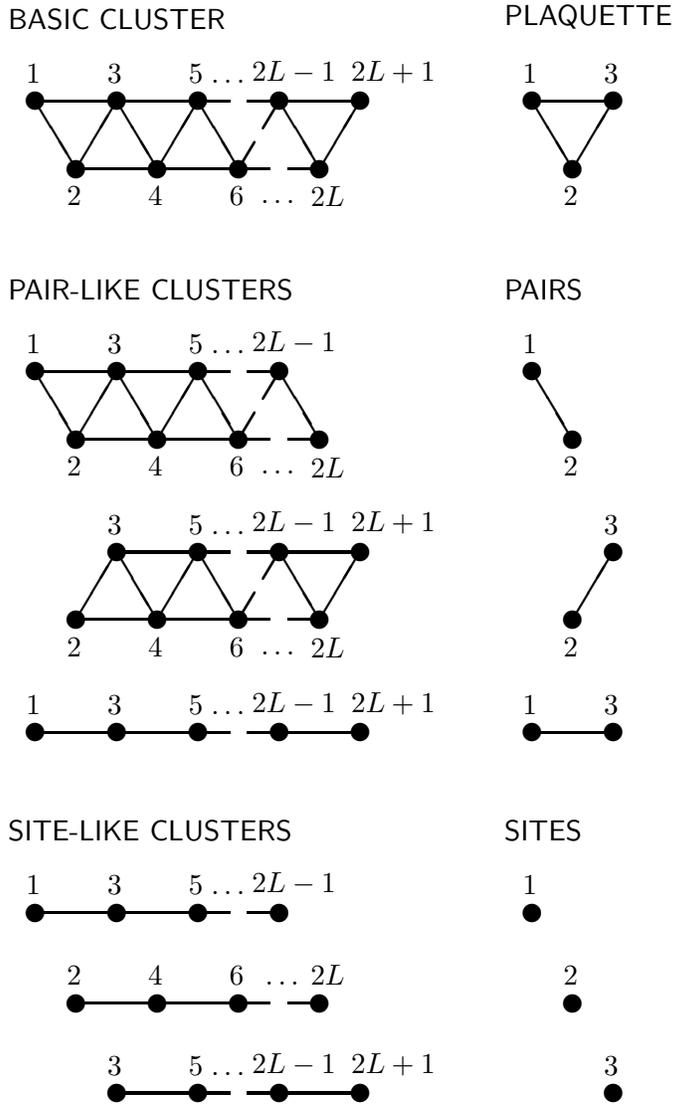

The B and C~hierarchies, originally proposed by Kikuchi and
Brush~\cite{KikuchiBrush1967}, are series of approximations with
increasing cluster size, suitable for 2d
square~\cite{KikuchiBrush1967} and
triangular~\cite{PelizzolaPretti1999} lattices. They are
interesting mainly because they converge towards the exact free
energy, in spite of the fact that the cluster size increases only
in one direction. This result has been proved rigorously only for
the C~hierarchy~\cite{Schlijper1983}, but there are numerical
evidences for both~\cite{KikuchiBrush1967,PelizzolaPretti1999}.
Such results~\cite{Schlijper1983} are related to the transfer
matrix concept: As the Bethe approximation solves exactly an
Ising-like chain, the CVM, with infinitely long 1d stripes as
basic clusters (to which the B and C~hierarchies tend), solves
exactly a 2d lattice. Here we are interested in showing that these
approximations verify the sufficient condition for the convergence
discussed above. Let us consider for instance the B~hierarchy on
the triangular lattice (a completely analogous treatment holds for
the C~hierarchy and/or for the square lattice). The basic
clusters, shown in Fig.~\ref{fig:gerb} (top row, left column), are
made up of a sequence of $L-1$~up- and $L$~down-pointing
triangles, where $L$ is an adjustable parameter. Of course, also
corresponding clusters with $L$~up- and $L-1$~down-pointing
triangles are allowed, but all basic clusters always extends only
in one direction. This choice can be viewed as a generalization of
the triangle plaquette approximation (see Fig.~\ref{fig:gerb}, top
row, right column), where of course also up-pointing triangles are
included in the set of basic clusters. In the following rows of
Fig.~\ref{fig:gerb} also the subclusters of the given basic
cluster, having nonzero coefficients in the cluster entropy
expansion ($a$-coefficients), are displayed. They are divided in
pair-like and site-like subclusters, in that they can be put in
one-to-one correspondence with pair and site subclusters for the
triangle plaquette approximations. Such analogy is not only a
pictorial one. In fact, it is possible to show (for instance
making use of Eq.~\eqref{eq:sumrule}, but see also
Ref.~\cite{KikuchiBrush1967}) that the $a$-coefficients are
$a_2=-1$ for pair-like clusters and $a_1=1$ for site-like
clusters, like for the triangle plaquette approximation. The same
holds for $c$-coefficients, i.e., the numbers of basic clusters
sharing a given subclusters, which turn out to be $c_2=2$ and
$c_1=6$ respectively, whence $b_2=-1/2$ and $b_1=1/6$. Finally,
from Fig.~\ref{fig:gerb} one easily sees that also the same
``allocation'' technique as for the plaquette approximation can be
used. Inside a given basic cluster, each site-like subcluster is
shared by $r=2$ pair-like clusters, and each pair-like cluster
contains 2 site-like subclusters, whence
inequality~\eqref{eq:condsuffplaq} is satisfied.

\subsection{Hypercube approximation in $d$ dimensions}

Finally, let us consider the case of a hypercubic lattice in
$d$~dimensions, and let us choose a $d$-dimensional hypercube
($d$-cube) as basic cluster. Of course, the relevant cases are
$d=2,3$, the former of which coincides with the square plaquette
approximation, mentioned above, but the interest of a general
treatment will be clearer later. It is possible to show, by
repeated use of Eq.~\eqref{eq:sumrule}, that clusters with non
zero coefficients are only $i$-cubes, for $i=1,\dots,d$, and the
$i$-cube coefficient in $d$ dimensions is $a_i^{(d)} =
(-1)^{d-i}$. Moreover, the number of $d$-cubes sharing a given
$i$-cube (in $d$ dimensions) is $c_i^{(d)} = 2^{d-i}$. As a
consequence, the normalized coefficients turn out to be
\begin{equation}
  b_i^{(d)} = \left( -\frac{1}{2} \right)^{d-i}
  .
  \label{eq:bcoeff_hcube}
\end{equation}
Let us now impose the sufficient conditions,
Eqs.~\eqref{eq:suffcondplus} and~\eqref{eq:suffcondminus}. Let us
notice that the positive coefficients, those who give problems for
upperbounding, have the $i$~index with the same parity as~$d$,
that is $i=d-2,d-4,\dots$. Then we can couple each $i$-cube with
$(i+1)$-cubes that contain it and are contained in a given
$d$-cube. As for plaquette approximations, let us split the
$i$-cube coefficient in equal parts, so that we have a single
$b_{i+1|i}^{(d)}$ allocation coefficient. We still have to observe
that each $i$-cube is shared by $d-i$ $(i+1)$-cubes contained in
the same $d$-cube (the equivalent of the $r$~parameter for
plaquette approximations), and that each $(i+1)$-cube contains
$2(i+1)$ different $i$-cubes (the equivalent of $2$~sites in a
pair). We can then rewrite Eqs.~\eqref{eq:suffcondplus}
and~\eqref{eq:suffcondminus} as
\begin{eqnarray}
  b_i^{(d)} & = & (d-i) \, b_{i+1|i}^{(d)}
  \\
  -b_{i+1}^{(d)} & \geq & 2(i+1) \, b_{i+1|i}^{(d)}
  .
\end{eqnarray}
By eliminating the allocation coefficient, we obtain
\begin{equation}
  \frac{b_i^{(d)}}{d-i} + \frac{b_{i+1}^{(d)}}{2(i+1)} \leq 0
  ,
  \label{eq:condsuffhcube}
\end{equation}
which, replacing Eq.~\eqref{eq:bcoeff_hcube} and taking into
account that $d-i$ is always even (as previously mentioned),
becomes
\begin{equation}
  2i \leq d-1
  .
\end{equation}
Such inequality becomes more and more difficult to be satisfied as
the subcluster index~$i$ increases. Therefore we have to consider
the worst case, that is $i=d-2$, leading to
\begin{equation}
  d \leq 3
  .
\end{equation}
This results essentially proves the convergence for $d=3$, because
the $d=2$ case coincides with the square plaquette approximation.
Nevertheless, it is mainly interesting in that it gives us the
opportunity to experiment the natural iteration method in a case
in which the sufficient condition is not verified. We have
actually implemented the procedure for the simple Ising model on
the $d=4$ hypercubic lattice, easily finding cases in which the
behavior is non convergent (oscillating). This fact lead us to
conjecture that actually the sufficient condition might be also a
necessary one.

\section{An equivalent formulation}

In a recent paper~\cite{HeskesAlbersKappen2003}, a general method
for the minimization of non convex functionals, related to the
existence of suitable upperbounds to the functional to be
minimized, is proposed and applied to the case of the CVM free
energy. Different possible choices for the upperbounding
functional are investigated. Hereafter, we show that one choice
proposed there, which by the way turns out to be quite convenient
in terms of computation time, is equivalent to the natural
iteration method. First, let us briefly recall the general method,
which is based on the following.

\noindent {\bf Theorem:} \ Let $F[p]$ be a continuous functional
in the set of variables $p$, defined in some compact
domain~$\Omega$, and $\bar{F}[p,p']$ an auxiliary continuous
functional in a pair of variable sets $p,p'$, defined in the
domain~$\Omega^2$, having a unique minimum with respect to~$p'$
for each fixed~$p$. Let the auxiliary functional satisfy the
following requirements:
\begin{eqnarray}
  &&
  F[p'] \leq \bar{F}[p,p']
  \label{eq:flefbar} \\ &&
  F[p'] = \bar{F}[p,p'] \ \ \Longleftrightarrow \ \ p' = p
  ,
  \label{eq:feqfbariff}
\end{eqnarray}
that is, the auxiliary functional is an upperbound to the original
functional, and equality holds if and only if the two arguments of
the former are equal. Then the application $\varphi: p \mapsto
\hat{p}$ defined by
\begin{equation}
  \hat{p} = \arg\min_{p' \in \Omega} \bar{F}[p,p']
  \label{eq:application}
\end{equation}
enjoys the properties
\begin{eqnarray}
  &&
  F[\hat{p}] \leq F[p]
  \label{eq:flef} \\ &&
  F[\hat{p}] = F[p] \ \ \Longleftrightarrow \ \ \hat{p} = p
  .
  \label{eq:feqfiff}
\end{eqnarray}
Therefore, it defines an iterative method to minimize the original
functional.

\noindent {\bf Proof:} \ It is easy to obtain the following
inequality chain
\begin{equation}
  F[\hat{p}] \leq \bar{F}[p,\hat{p}] \leq \bar{F}[p,p] = F[p]
  ,
  \label{eq:ineqchain}
\end{equation}
proving immediately Eq.~\eqref{eq:flef}. The first inequality is
the first hypothesis on the auxiliary functional~$\bar{F}$,
Eq.~\eqref{eq:flefbar}; the second inequality is a consequence of
the definition of~$\varphi$, Eq.~\eqref{eq:application}; the
equality descends from the second hypothesis on~$\bar{F}$,
Eq.~\eqref{eq:feqfbariff}. In order to prove also
Eq.~\eqref{eq:feqfiff}, we have to show that both inequalities
hold as equalities if and only if~$\hat{p} = p$. As far as the
former is concerned, this is a direct consequence of the
hypothesis Eq.~\eqref{eq:feqfbariff}, while the latter is proved
by the fact that~$\bar{F}[p,p']$ has a unique minimum, which is
also the absolute minimum, with respect to~$p'$.~$\blacksquare$

Let us now consider the auxiliary functional defined by
\begin{equation}
  \bar{F}[p,p'] =
  \sum_{\gamma}
  \sum_{x_\gamma} p'_\gamma(x_\gamma)
  \left[
  h_\gamma(x_\gamma)
  + \log p'_\gamma(x_\gamma)
  + \sum_{\alpha \subset \gamma} b_\alpha \log p_\gamma(x_\alpha)
  \right]
  .
\end{equation}
First of all, it is easy to see that $\bar{F}[p,p] = F[p]$, where
$F[p]$ is the CVM free energy~\eqref{eq:f1}. Moreover, $F[p,p']$
is easily seen to be convex with respect to~$p'$, therefore, if it
has a stationary point, it is also unique, and is a minimum.
Finally, let us observe that stationarization of this functional
with respect to~$p'$, with the usual linear constraints, gives
rise just to the NIM equations~\eqref{eq:nim}, which in this way
can be used to define the application~$\varphi$. In order to show
that $\varphi$ actually perform a minimization of~$F$, a
sufficient condition is given by
Eqs.~\eqref{eq:flefbar},\eqref{eq:feqfbariff} in the above
theorem, that is, we have to upperbound the quantity
\begin{equation}
  F[p'] - \bar{F}[p,p'] =
  - \sum_\gamma \sum_{x_\gamma}
  p'_\gamma(x_\gamma)
  \sum_{\alpha \subset \gamma}
  b_\alpha \log \frac{p_\gamma(x_\alpha)}{p'_\gamma(x_\alpha)}
  \label{eq:deltaf4}
\end{equation}
with zero. Going back to (the right hand side of)
Eq.~\eqref{eq:deltaf3}, it easily turns out that this is exactly
the same upperbound we have proved with the sufficient condition
for the convergence of the NIM.

\section{Conclusions}

Let us finally summarize our results. We have investigated on the
convergence of the natural iteration method, proposed by Kikuchi
as a minimization procedure for cluster variational free energies
and widely employed in a lot of applications of the CVM. We have
discussed a condition on the coefficients of the cluster entropy
expansion, which is sufficient to prove that the free energy
decreases at each iteration, ensuring the convergence of the
method. Such a condition is based on the idea of pairing
subcluster entropies with a positive coefficient to larger
subcluster terms with a negative coefficient, yielding a set of
conditional entropy terms with negative coefficients. It had
already been proved by Kikuchi in the original
paper~\cite{Kikuchi1974} that negative coefficient terms give
decreasing contributions to the free energy. We have also taken
into account a set of common CVM approximations defined on various
regular lattices, frequently encountered in applications, showing
that the sufficient condition is always satisfied. In particular,
we have devoted some attention to the class of hypercube
approximations on the generic ($d$-dimensional) hypercubic
lattice, showing that the sufficient condition is verified for $d
\leq 3$. We have also implemented the natural iteration method for
$d=4$ on the simple Ising model, and found out that several
(random as well as uniform) initial conditions give rise to non
convergent (oscillating) behavior. This fact has led us to
conjecture that the sufficient condition may be also a necessary
one. Finally we have established a connection with a recently
proposed method for the minimization of non-convex functionals,
which can be applied to the CVM free
energy~\cite{HeskesAlbersKappen2003}. Such a method is based on
the existence of suitable upperbounding functionals to the
functional to be minimized. In Ref.~\cite{HeskesAlbersKappen2003}
several choices of upperbounding functionals are proposed and
applied to simple inhomogeneous systems. We have shown that one of
the upperbounding choices proposed there (indeed quite a good
choice in terms of computation time) is actually equivalent to
Kikuchi's natural iteration method. It turns out explicitly that
the upperbounding condition implies free energy decreasing, whence
convergence.

\begin{acknowledgments}
I would like to express my thanks to Dr. Alessandro Pelizzola for
many helpful suggestions and discussions.
\end{acknowledgments}


\clearpage

\begin{table}[p]
  \caption{
    Coefficients for different plaquette approximations. The first
    two columns report respectively the lattice and plaquette (basic cluster) type.
    The following three columns display the independent
    coefficients: $q$ (coordination number), $c_2,c_1$ (number of
    plaquettes sharing a given pair, site). The last two columns
    verify the sufficient condition, in that $q(c_2-1) < 2(c_1-1)$.
  }
  \begin{ruledtabular}
  \begin{tabular}{ll|rrr|rr}
    lattice    & plaquette   & $q$ & $c_2$ & $c_1$ & $q(c_2-1)$ & $2(c_1-1)$ \cr
    \hline
    square     & square      &  4 & 2 &  4 &  4 &  6 \cr
    triangular & triangle    &  6 & 2 &  6 &  6 & 10 \cr
    honeycomb  & hexagon     &  3 & 2 &  3 &  3 &  4 \cr
    sc         & square      &  6 & 4 & 12 & 18 & 22 \cr
    fcc        & triangle    & 12 & 4 & 24 & 36 & 46 \cr
    fcc        & tetrahedron & 12 & 2 &  8 & 12 & 14
  \end{tabular}
  \end{ruledtabular}
  \label{tab:coefficients}
\end{table}

\clearpage

\end{document}